\documentclass[epj,final]{svjour}
\pdfoutput=1

\usepackage[applemac]{inputenc}
\usepackage{graphicx}
\usepackage{amsmath}
\usepackage{amssymb}
\usepackage{times}
\usepackage{verbatim}
\usepackage{booktabs}
\usepackage{url}

\DeclareGraphicsExtensions{.pdf}

\begin{document}




\title{The Swiss Board Directors Network in 2009}



\author{Fabio Daolio\inst{1} \and Marco Tomassini\inst{1}\and Konstantin Bitkov\inst{1}}

\institute{Faculty of Business and Economics, University of Lausanne, Switzerland}

\date{Submitted: March 10, 2011\thanks{With kind permission of The European Physical Journal (EPJ)}}

\abstract{
\noindent We study the networks formed by the directors of the most important Swiss boards and the
boards themselves for the year 2009. The networks are obtained by projection from the original bipartite graph.
We highlight a number of important statistical features of those networks such as degree distribution, weight
distribution, and several centrality measures as well as their interrelationships. While similar statistics were already known
for other board systems, and are comparable here, we have extended the study with a careful investigation of director and board centrality,
 a k-core analysis, and a simulation of the speed of
information propagation and its relationships with the topological aspects of the network such as clustering and
link weight and betweenness. The overall picture that emerges is one in which the topological structure of
the Swiss board and director networks has evolved in such a way that special actors and links between actors play a
fundamental role in the flow of information among distant parts of the network. This is shown in particular by
the centrality measures and by the simulation of a simple epidemic process on the directors network.
\PACS{
      {89.65.Gh}{Economics; econophysics, financial markets, business and management}   \and
      {89.75.-k}{Complex systems}   \and
      { 89.75.Fb}{Structures and organization in complex systems}   \and
      {87.23.Ge}{Dynamics of social systems}
     } 
\keywords {Boards of directors -- Complex networks -- Corporate governance}
} 

\maketitle

%


\section{Introduction}
\label{intro}

Corporate governance as expressed by directors boards plays a fundamental role in the economy
of a country and, through multinational firms, the influence may also reach other countries. Given that many board directors
usually sit on more than one board, a web of relationships between boards and directors implicitly arises.
These affiliation networks are potentially a useful tool to understand information flow and influence between
companies. In fact, board decisions and practices 
may diffuse and percolate through the network and knowledge of the structure of the latter becomes of the utmost interest
if one wants to understand the dynamics of such phenomena.
Through the use of well established techniques in complex network theory~\cite{guido,newman-book} it is now possible to study the structure
of corporate boards networks in great detail. Indeed, a few studies dealing with the subject have been published recently.
In particular, we mention Davis' et al. work on the American corporate \'elite~\cite{davisUS}, and a couple of similar 
investigations dealing with the Italian and American corporate board systems~\cite{calda1,battiston1}. Although the details do
differ, it turns out that there are several strikingly common features across different countries and over a span of time. 
Of course, there can be many other conceivable ways in which board directors may interact outside of the board meetings, such
as shared service on educational, non-profit, and even belonging to the same country club. However, all these possible
connections are very difficult to disentangle and quantify, in contrast with straight affiliation to a board. It is thus very likely that
the composition of corporate boards are a sufficiently good proxy.

In this paper we present an investigation along the same lines of the Swiss corporate boards network. 
This case study is an interesting one for several reasons as Switzerland plays an important role in corporate finance
and in other production sectors throughout the world. 
A recent research paper has dealt with the interesting issue of gender diversity and nationality in Swiss corporate
boards~\cite{ruigrok}.
However, a study of the structural aspects of Swiss boards from the point of view of complex networks is still
missing, to the best of our
knowledge, except for~\cite{ginalski} which, however, puts the emphasis on the sociology of industrial family networks
in the country during the 20th century and uses time-resolved data up to the year 2000 only. Here, on the other hand, we
focus on several important network characteristics using data from the year 2009, which is interesting since they 
belong to a period that immediately follows the onset of
the recent world-wide financial and economical crisis.  
We shall study in detail the structure of the Swiss boards directors network in order to pave the
way for a better understanding of decision making processes and how the web of relationships between firms and
board directors may influence it. However, we shall limit ourselves to the general inferences that can reasonably be made
on structural considerations alone, refraining from attempting to provide sociological or managerial analyses
of corporate strategy and practices. Such a sociological analysis, taking into account
historical evolution and managing practices as well as network structure can be found, for example, 
in Davis et al. in the US case~\cite{davisUS}.

\begin{figure*}[htb!]
\begin{center}
 \includegraphics[width=0.3\textwidth]{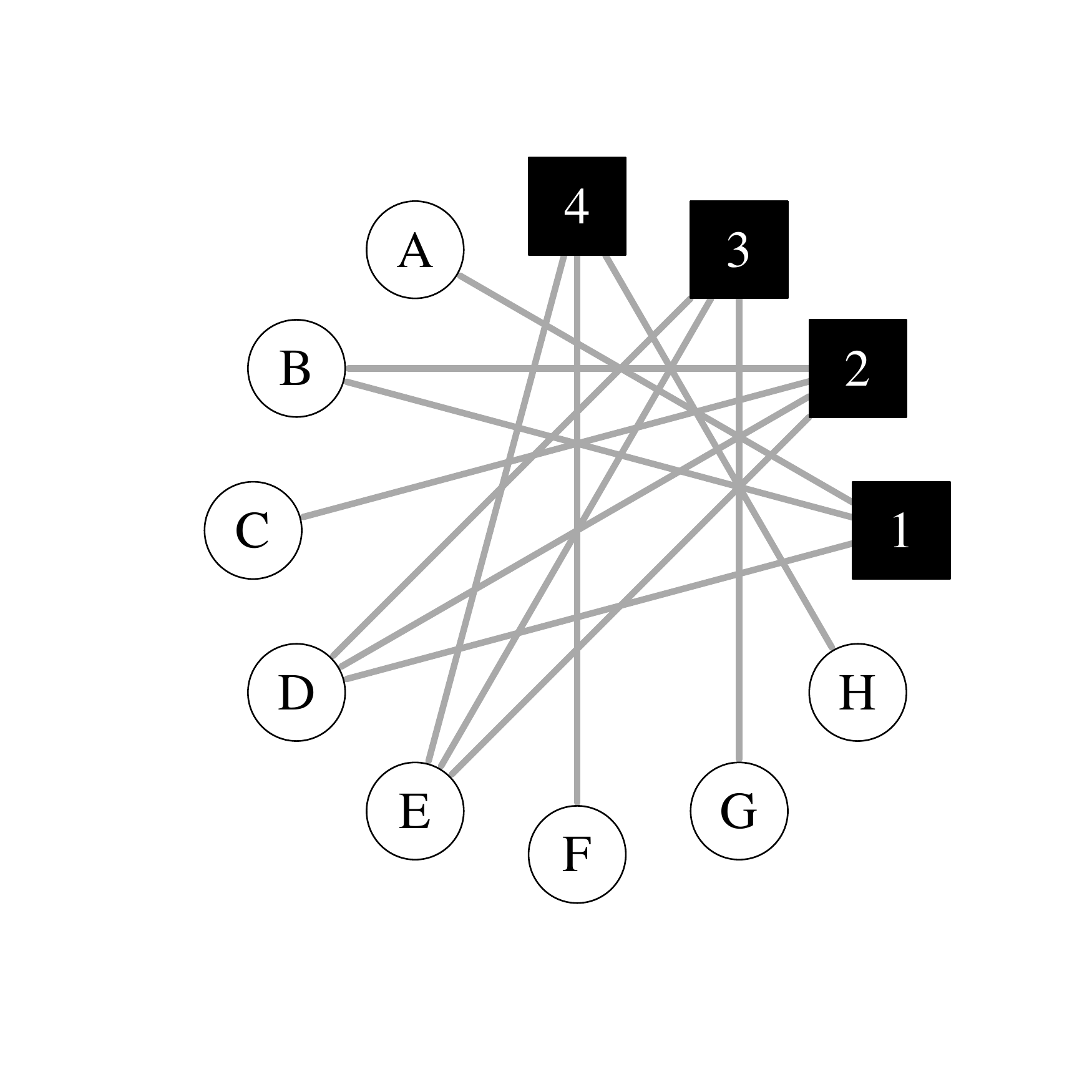}
 \includegraphics[width=0.28\textwidth]{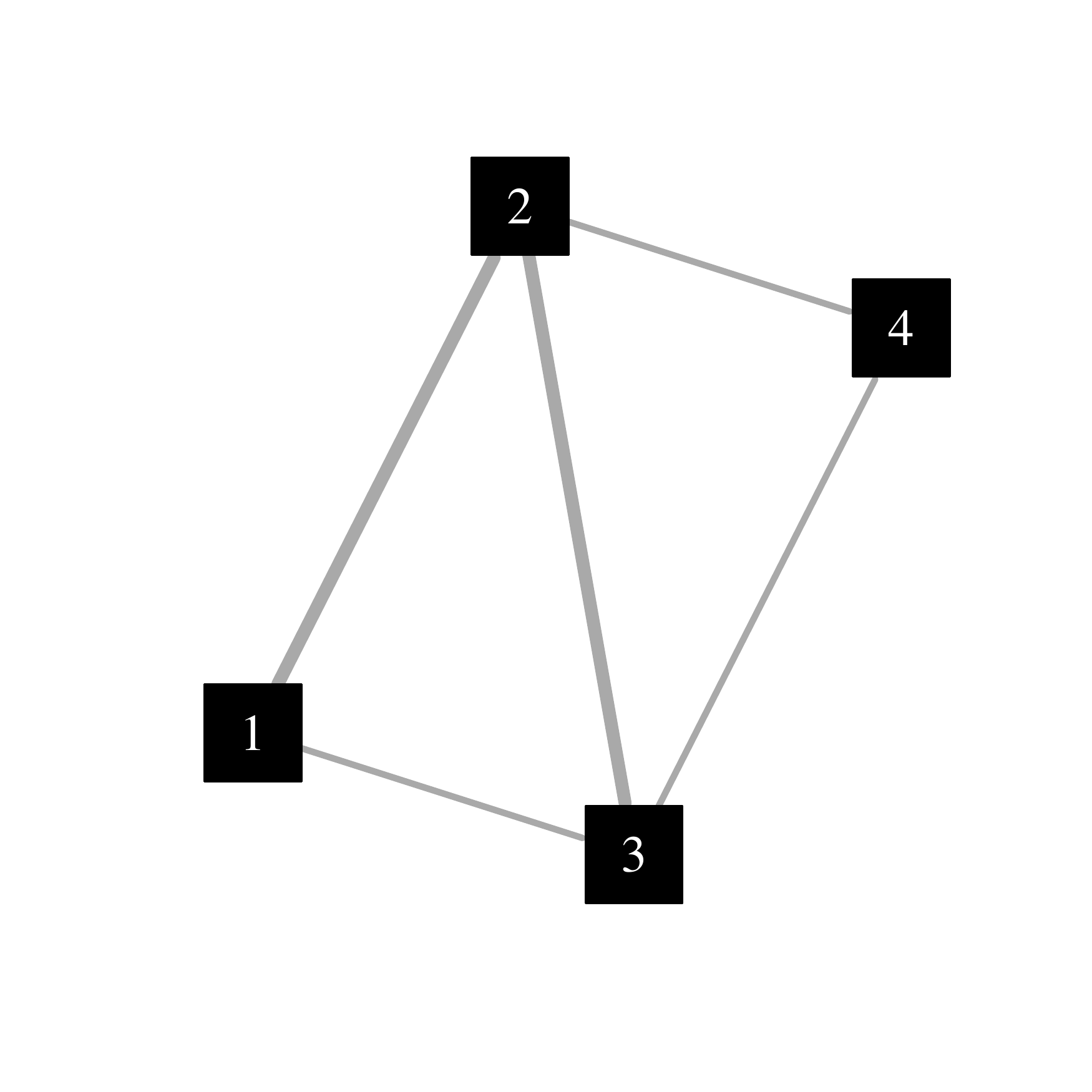}
 \hspace{-20pt}
 \includegraphics[width=0.28\textwidth]{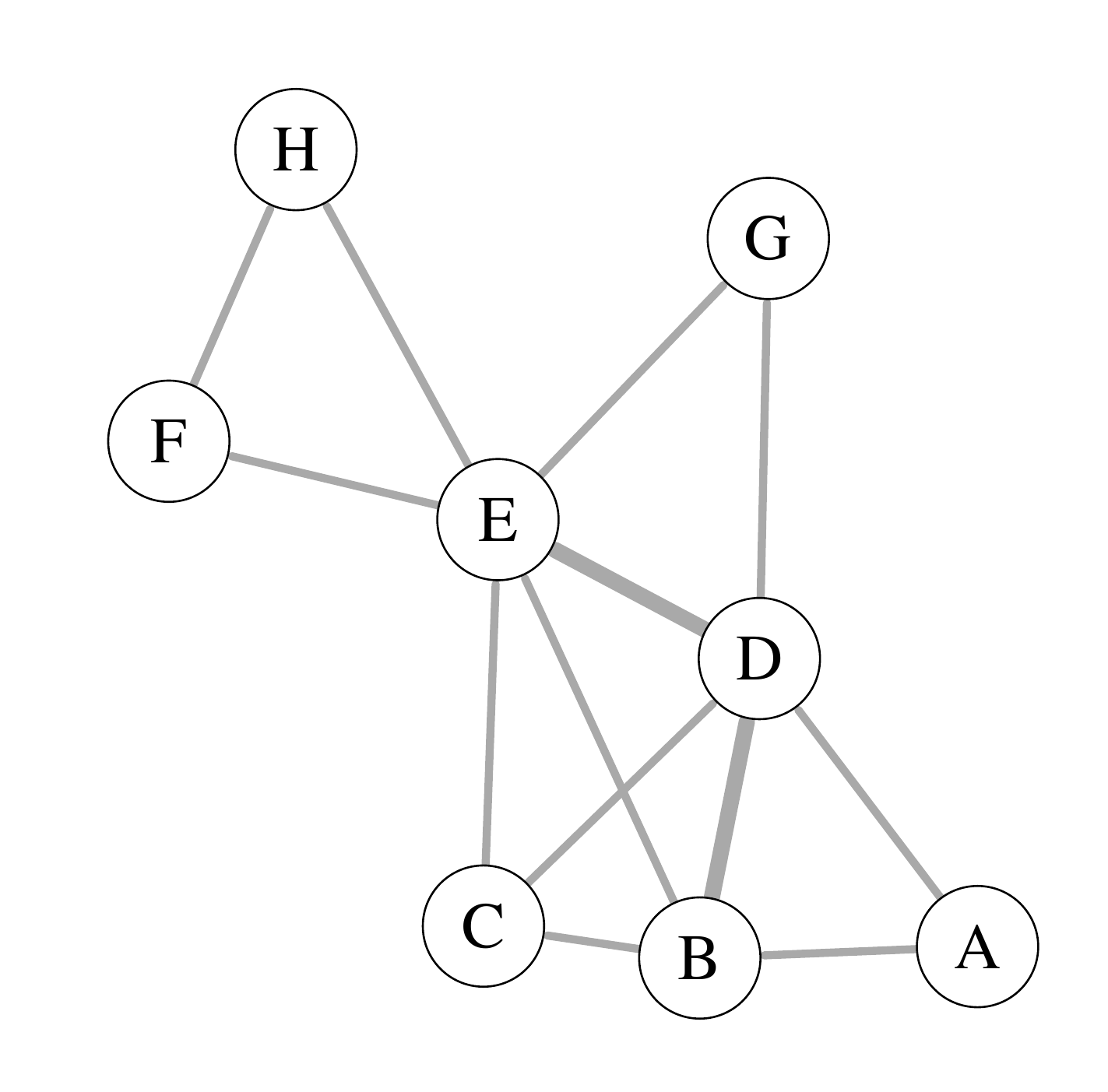}
 \caption{A small example of a bipartite graph. The leftmost image is the original bipartite graph in which circles
 represent directors nodes and squares stand for company boards. The middle image is the boards projection
 graph and the rightmost image is the directors projection. The thickness of the links is proportional to the
 corresponding edge weight.}
 \label{bip}
\end{center}
\end{figure*}

The paper is organized as follows: in the next section we describe the data collection process and the construction of
the network from the raw data. We then present the main network statistics and their relationships with the social and
economic entities they represent. This is followed by a simple analysis of the propagation of ideas, decisions, or influences
through the network and of the role of tie strength and tie position. Finally, we give our conclusions.

\section{Network Construction and Structure}

In this section we present the basic concepts for the directors and boards network construction and description starting from the row data
in order to make the article self-contained. Much more detailed explanations can be found in standard texts on complex networks
such as~\cite{guido,newman-book}. 

\subsection{Data}

We have collected the data from publicly available sources, essentially annual reports of companies, and their web sites. Board structure
in Switzerland is based in the majority of cases on a two-tier system; this means that there is a supervisory board of directors and
a second separate management board of executive directors that meet separately. This is also the case in Germany and Austria
 for example, but some
other countries have only either a single board or a mixed system where the two boards meet separately, but some executive
directors sit on the supervisory board. We have collected data on both boards for companies that adhere to the
two-tier system but, for simplicity and for the
sake of comparison with countries where the system is different, in this work we only use the supervisory boards data.

The sample consists
of the $108$ top revenue companies in Switzerland with a total of $818$ distinct directors in 2009. The ranking we used is based on a $2005$ 
report~\cite{top-magazine} and it has been integrated with the information available on relevant magazines and websites~\cite{forbes,money,help} for an update. This should not produce any noticeable bias in the selection, as the Swiss corporate landscape has been dominated by a few tens of
big companies, only a small number of which have changed by acquisition, bankruptcy, or mergers in the last ten years. Then, the manually collected data on the boards of those companies, have been
carefully checked for names that are reported differently in different boards but correspond to the same person, and also true 
duplicates. In the collection process we have also recorded, when available, data on age, gender, and nationality
of directors.


\begin{table*}[ht!]
\begin{center}
\caption{Average and global quantities for board networks (B) and director networks (D)\label{avg} of the top companies in Switzerland (CH,2009), Italy (IT,2002), and United States of America (US,1999).
$N$=number of nodes, $E$=number of edges, $N_c/N$=relative size of the largest connected component, $\langle k \rangle$=average degree, $\langle k \rangle /k_c$=network density, $\overline{C}$ =average clustering coefficient, $d$=average path length.} 
\begin{tabular}{ccccccccc} \toprule
\multicolumn{1}{}{}&\multicolumn{1}{c}{B-CH,09}&\multicolumn{1}{c}{D-CH,09}&\multicolumn{1}{c}{B-IT,02}&\multicolumn{1}{c}{D-IT,02}&\multicolumn{1}{c}{B-US,99}&\multicolumn{1}{c}{D-US,99}\\ \cmidrule{2-7}
$N$ 						&108& 818& 240&1906&916&7680\\ 
$E$						&91& 3971& 636&12815&3321&55437\\ 
$N_c/N$					&0.62 &0.61&0.82&0.84&0.87&0.89\\ 
$\langle k \rangle$			&1.69 &9.71&5.30 & 13.45 &7.25 &14.44\\
$\langle k \rangle /k_c(\%)$	& 1.57&1.19& 2.22&0.71&1.57&0.79\\ 
$\overline{C}$				& 0.246& 0.859& 0.318&0.915&0.376&0.884\\ 
$d$ 						& 6.4& 7.2& 4.4&3.6&4.6&3.7\\ 
\bottomrule
\end{tabular}
\end{center}
\end{table*}

\subsection{Bipartite Graphs and Projections}

Once the directors in each board are know, one can obtain a network by assigning a node to each director and to each
board.  Going through the directors nodes and  tracing an edge between a given director and the boards he/she sits in,
produces a network that is called a \textit{bipartite graph}.
A graph $G(V,E)$ in which $V=\{v_1, \ldots, v_N\}$ is the set of vertices or nodes, and $E=\{e_1, \ldots, e_M\}$ is the set
of edges or links, is said to be bipartite when the vertices can be partitioned into two disjoint sets $V =V_1 \cup V_2$,
$V_1 \cap V_2 = \emptyset$,
such that there are no edges $e=\{u,v\}$ between vertices belonging to the same set: 
$$ \{\{u,v\} : u \in V_1, v \in V_2\}, \; \forall e \in E.$$

This can be depicted as in the leftmost image of Fig.~\ref{bip} where a set of nodes (circles) represents directors 
and the other set (squares) represents the boards.
A link between a director and a board means that the director sits in that board. When two boards share the same director 
it is said that there is an \textit{interlock}. \textit{Multiple interlocks} are also possible, in which at least two directors of a board
sit together on another board. The \textit{incidence matrix} $B$ of a bipartite network with, say, $l$ boards and $m$ directors
is an $l \times m$ rectangular matrix such that the generic matrix element $B_{ij}$ is $1$ if director $j$ belongs to board
$i$ and $0$ otherwise~\cite{newman-book}.

From the bipartite graph, it is an easy matter to obtain two derived graphs which are called
\textit{projections}. One can construct a graph in which two directors are connected if they sit on the same
board. Or we can also build the projection in which two boards are connected if they share a common director.
These two projections are schematically depicted in the central and right images of Fig.~\ref{bip}. The two projections capture the essence of the
relationships we are looking for but they do not account for the ``weight'' of a relationship. Indeed, it is sensible to
say that it is not the same whether two people sit together on a single board or on several. In some sense,
their degree of interaction should be higher in the latter case. To account for this, the projection can be weighted;
for example, for the directors projection, an edge, i.e. a pair of connected directors, will have a weight equal to the number 
of common boards. The weighted projection can be obtained from the incidence matrix $B$ as follows~\cite{newman-book}:
\begin{equation}
P = B^{T}B, \;\;\mbox{where}\;\; P_{ij} = \sum_{k=1}^{l} B_{ik}^{T}B_{kj}
\label{project}
\end{equation}
where $B^{T}$ is the transpose of $B$, and $l$ is the number of boards. The elements $P_{ij}$ of the $m \times m$ matrix $P$ are 
the weights, i.e. the number of common boards shared by directors $i$ and $j$, whereas the diagonal elements $P_{ii}$ are the
number of boards in which director $i$ sits.

\section{Statistical Analysis}

For the  statistical analysis, symbols have the following meanings. A network is a graph $G(V,E)$, where the
set of vertices $V$ represents the agents, and the set of weighted edges $E$ represents their symmetric interactions. 
Here, depending on which projection is studied, the vertices are directors or firms respectively.

\noindent The weight of an edge $e \in E$ will be denoted by $w_e$ or by $w_{ij}$, by using the edge endpoints $i$ and $j$.
The size of a graph $G$ is the number of edges $|E|$ but here we shall follow the physicists' convention and define the size
$N$ as the cardinality of $V$. 

\noindent A neighbor  of an agent $i$ is any other agent $j$ at distance one
from $i$ (ignoring the weight of the link). The set of neighbors of $i$  is called  $V_i$  and its cardinality is the degree $k_i$ of vertex $i \in V$. The average degree of the network is called $\langle k \rangle$.

\noindent Other important quantities based on the previous definitions will be introduced in the sequel as needed. All the computations
have been carried out with the package \textit{igraph}~\cite{igraph} in the statistical computing environment  \textit{R}~\cite{R}.

\subsection{Average Quantities}
\label{averages}

Table~\ref{avg} summarizes the results for the Swiss boards case and compares them with those found in previous
studies~\cite{battiston1,calda1} for the Italian and American cases in the years 2002 and 1999 respectively. These data
are reported here only for the sake of comparison as they refer to separated moments in time.  No doubt, the Italian
and American networks have evolved somehow in the meantime, but we are not aware of more recent results.

Except for the noticeably smaller average degree and size of the largest connected component, the global statistics of the boards and
director graphs are comparable with those of the American and Italian cases. The reason for smaller largest connected
components is related to the number of interlocks. The larger this number in the bipartite graph, the more connected the two projections. This
points to the fact that the interlock phenomenon is less acute in Swiss boards. 
The lower average degree $\langle k \rangle$ indicates that directors and boards alike are less densely connected in
the Swiss case and, in the same manner, the slightly larger
mean path lengths can also be attributed to a lower degree of interlock.

Although the present study is not focused on sociological issues, since we collected the corresponding data, it might be
interesting to note that as per gender diversity in Swiss boards in 2009 the percentage of women is $8.6\%$, which
agrees with the figure published in the 2009 report~\cite{heidrick-report}. There is an increasing trend since 2003 data had a $3\%$ fraction~\cite{ruigrok}.

Another piece of information is the average age of directors, which is slightly under $60$, in agreement with the
smaller sample ($19$ companies) used in~\cite{heidrick-report}. Finally, the proportion of non-national directors in
our sample turns out to be about $0.46$, which again agrees with the figures reported in~\cite{heidrick-report} and
represents a marked increase with the 2003 figure of $22.1\%$~\cite{ruigrok}. 
However, a caveat is in order here: the nationality of some directors was not available and a presumably small but unknown amount of 
directors are binationals.
Anyway, a new trend towards internationalization has clearly established itself in the last
few years, probably as a result of more advanced corporate practices and more transparency in difficult economic times. The
trend is common to several other European countries as well~\cite{heidrick-report}.

\begin{figure}[!ht] \center
 \includegraphics[width=0.4\textwidth]{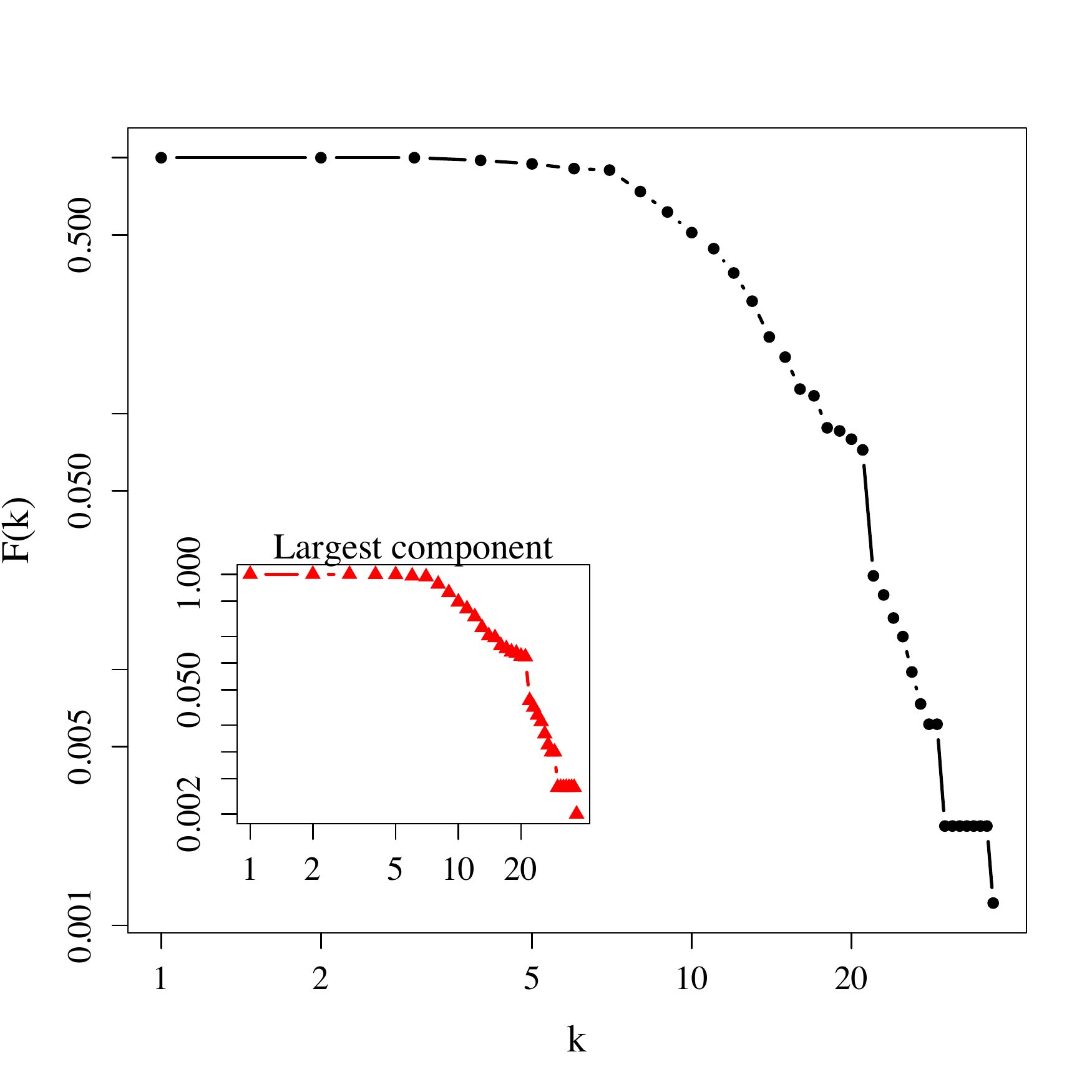}
 \includegraphics[width=0.4\textwidth]{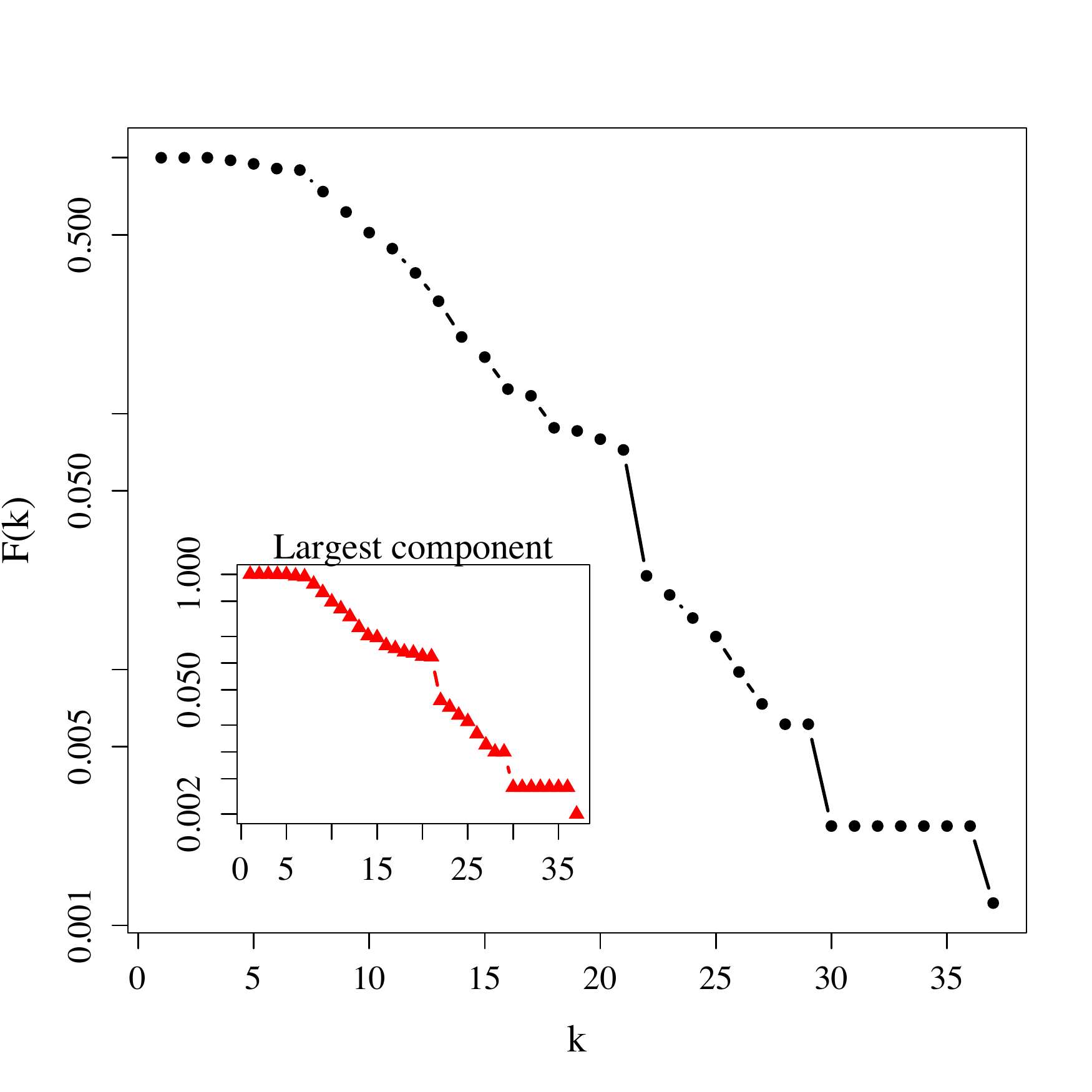}
 \caption{Directors network: empirical cumulative distribution function of degree, on log-log scale (top) and on lin-log scale (bottom). Large pictures refer to the whole network, the insets refer to the largest connected component.}
 \label{dcddf}
\end{figure}

\subsection{Distributions}

The degree distribution function $p(k)$ of a graph represents the probability
that an arbitrarily chosen node has degree $k$ or, equivalently,  the fraction of nodes with degree $k$.  Figure~\ref{dcddf} depicts
 the empirical cumulative degree distribution function $F(k)$, i.e. $F(k)=\sum^{k_{max}}_{k=k_{min}} p(k)$,
 for the directors network.

From the figure, it can be seen that the distribution falls off faster than a power-law, which is clear  from the log-log plot
on the top image of the figure.  In fact, the distribution decreases faster than linear on this plot; rather, it seems to be closer to an 
exponential, as seen from the bottom lin-log plot. 
However, owing to lack of sufficient data we refrain from trying to fit an analytical curve to the observed points. 
The insets show the corresponding distributions for the largest connected component only. It is clear that the distributions
are very similar to those referring to the whole graph.
On the one hand, these results differ
somewhat from those found for the Italian and American boards of directors, where the authors could observe a
power-law tail~\cite{battiston1,calda1}. On the other hand, we also observe a characteristic plateau in the distribution
at about $k=8$, which corresponds to the mean number of directors per board. In the case of ~\cite{battiston1,calda1} the average was about $10$. 

\begin{figure}[!ht]\center
 \includegraphics[width=0.4\textwidth]{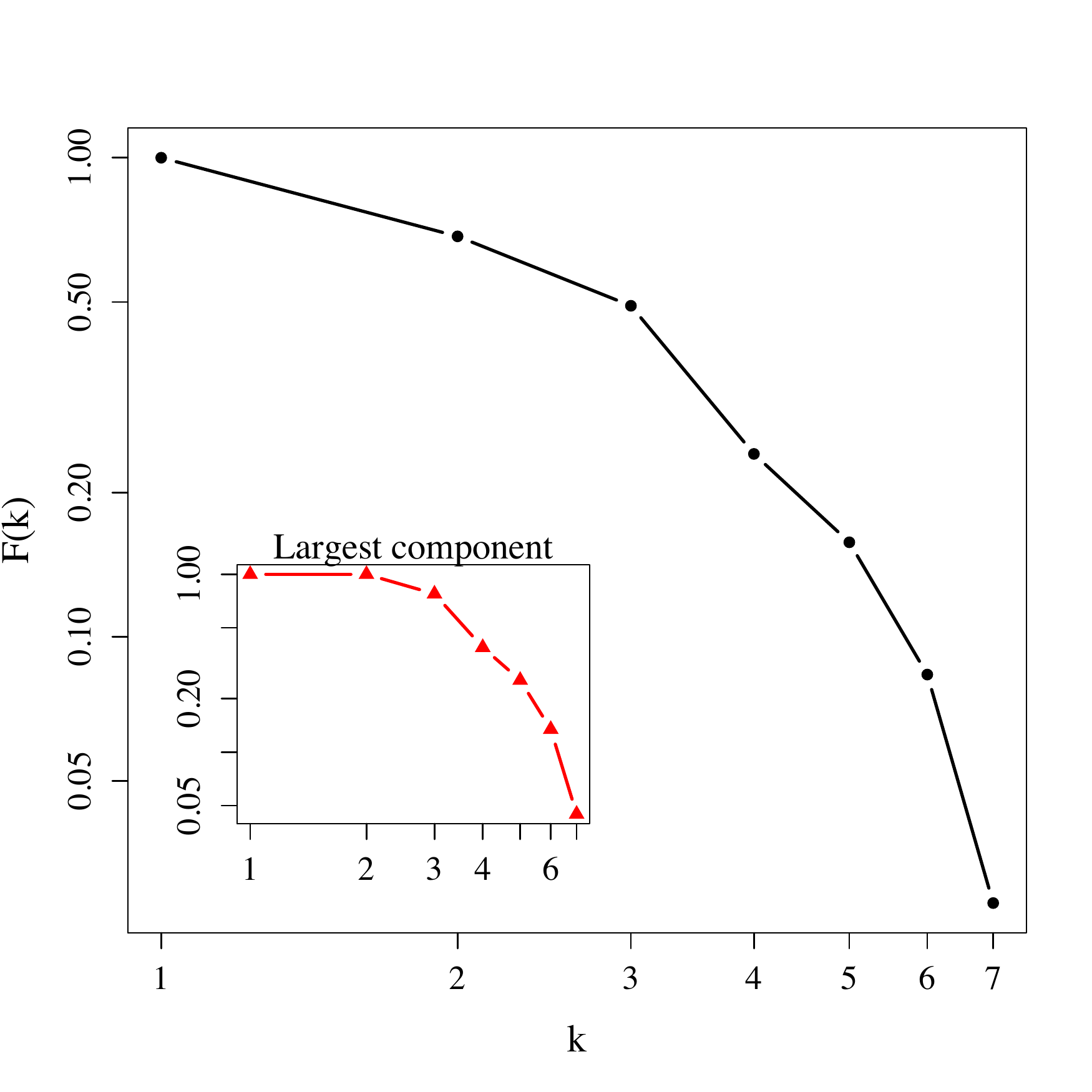}
 \includegraphics[width=0.4\textwidth]{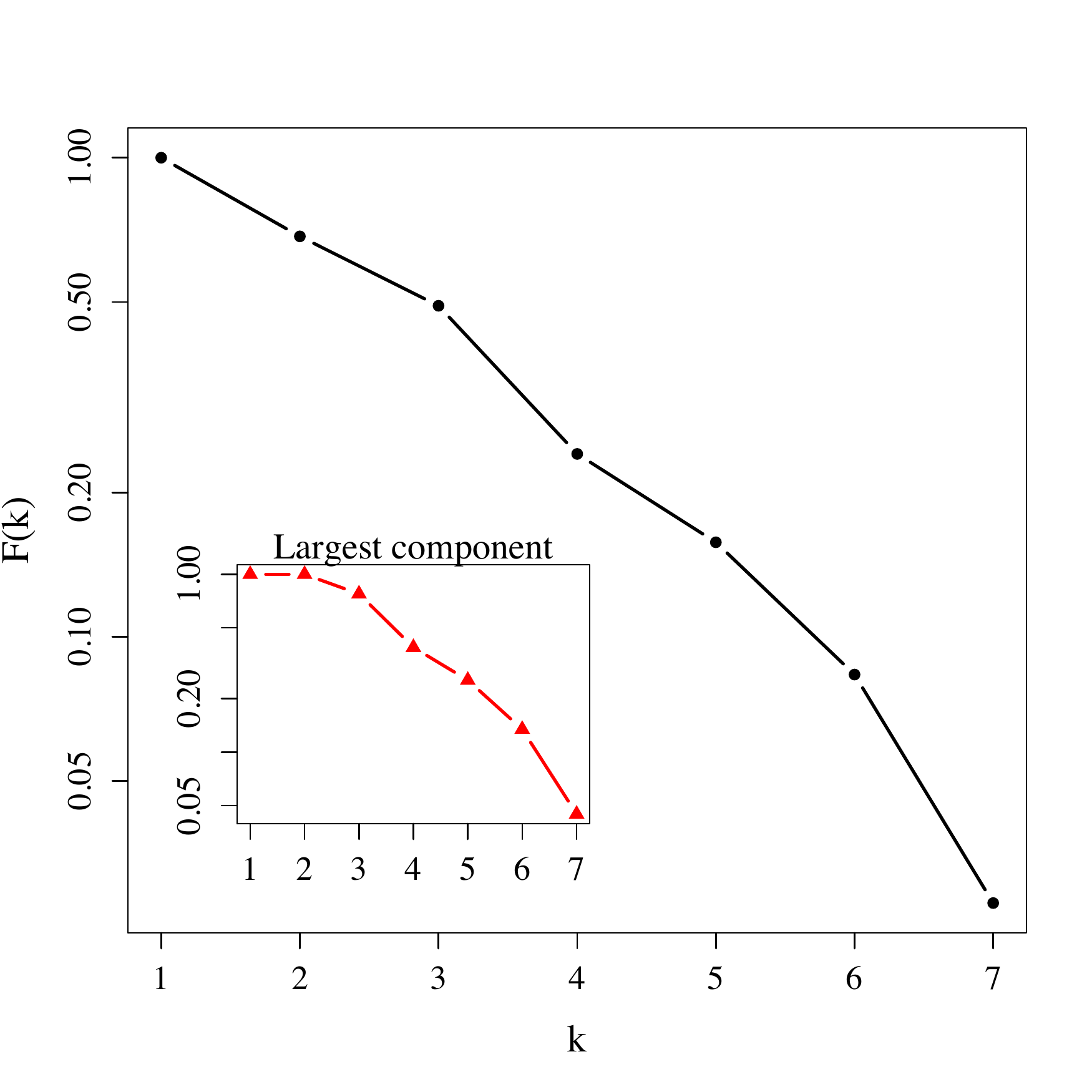}
 \caption{Boards network: empirical cumulative distribution function of degree, on log-log scale (top) and on lin-log scale (bottom). Large pictures refer to the whole network, the insets refer to the largest connected component.}
 \label{Bcddf}
\end{figure}

The board projection degree distributions, shown in Fig.~\ref{Bcddf}, do not present any notable feature. Owing to the small size of the network, the curves fall off quickly due to the limited degree range, as no degree larger than $7$ is present. The relatively poor connectivity of
boards confirms and reinforces the previous observation (Sect.~\ref{averages}) about the lower number of interlocks in the Swiss boards directors network.

The previous distributions were concerned with the topological aspects only; we present next two distributions that take into account the weighted 
nature of the networks that were obtained according to equation~\ref{project}. The link weight distribution is called $p(w_e)$ and
gives the probability that a randomly drawn link $e \in E$ has weight $w_e$. An analogous of the node degree for weighted networks
is the node \textit{strength} $s_i$ of a vertex $i \in V$ defined as $s_i = \sum_{j \in V_i} w_{ij}$,
i.e. the sum of the weights of the links incident in $i$~\cite{bart05}. The strength distribution 
$p(s)$ denotes the frequency of a given strength $s$ in the network. 

\begin{figure}[!ht] \center
 \includegraphics[width=0.4\textwidth]{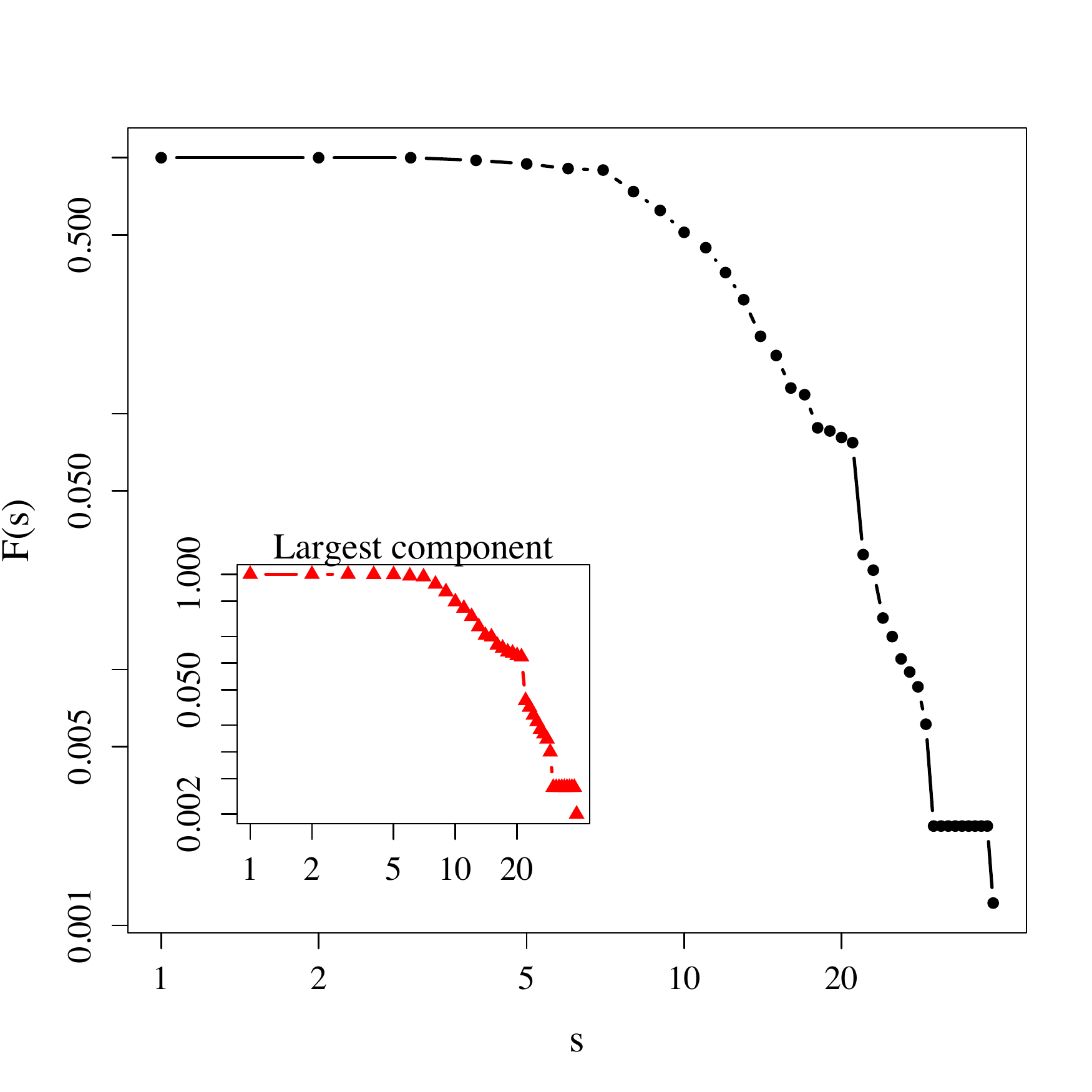}
 \includegraphics[width=0.4\textwidth]{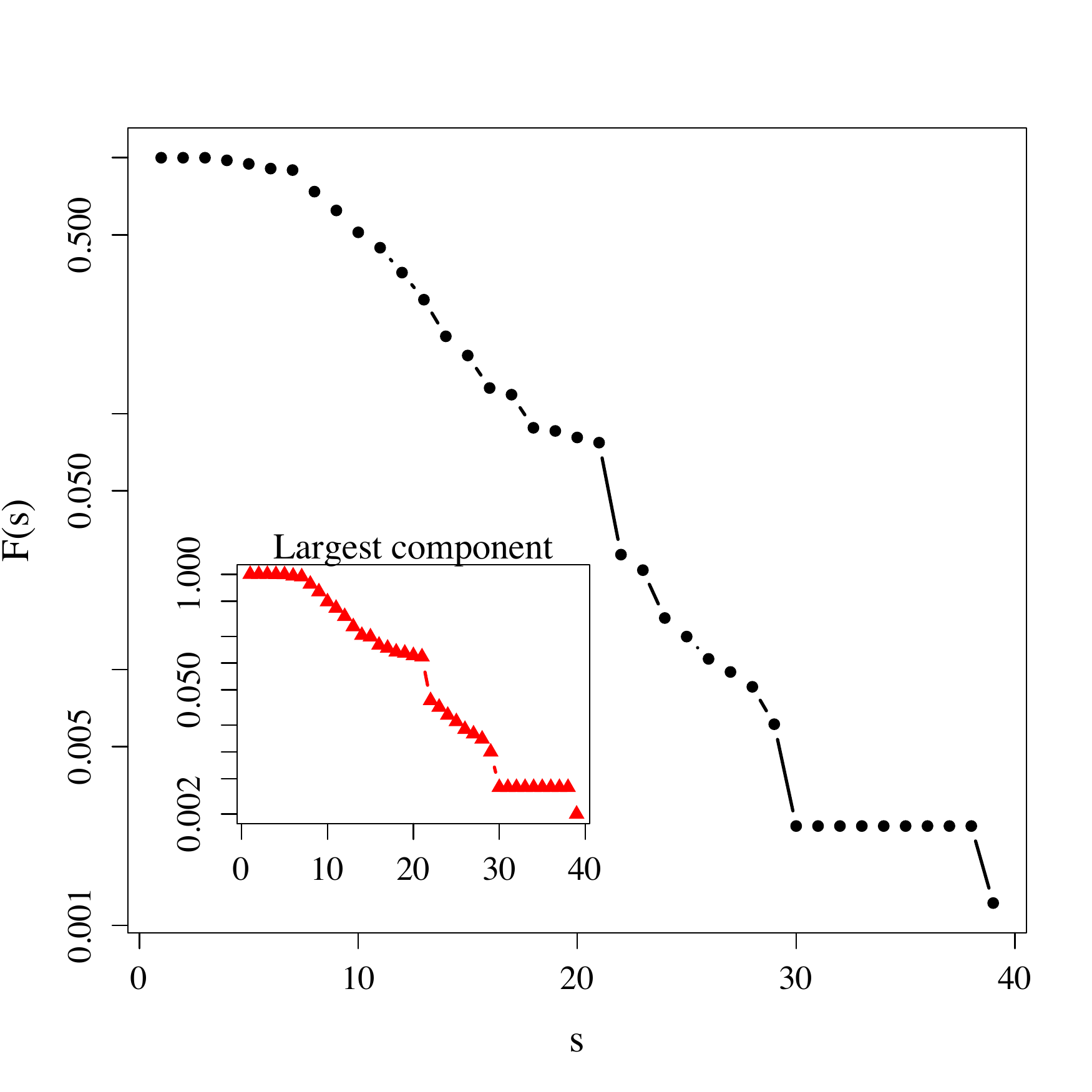}
 \caption{Directors network: empirical cumulative distribution function of the strength, on log-log scale (top) and on lin-log scale (bottom). Large pictures refer to the whole network, the insets refer to the largest connected component.}
 \label{strength-distr}
\end{figure}

In the present case, the $p(w_e)$ distributions are not
very informative and therefore
are not shown. In fact, in the directors projection, only $21$ out of $3971$ links have $w_e=2$; no weight greater then $2$ is found and all the remaining connections have unitary weight.
In the boards network, indeed, one can find values up to $3$ and $4$, but those are just two cases of companies belonging to same groups: ``Migros'' and ``Denner'' share $3$ directors, whereas ``Alpiq'' and ``EOS Holding'' share $4$.  Beside these, only $5$ out of $91$ links have a weight greater than one. This feature points to the fact
that multiple interlocks are extremely rare in the Swiss case; this is not the same as the American and especially the Italian examples, 
where one observes weights up to $6$ and a longer tail of the distribution. 

Figure~\ref{strength-distr} is a plot of the cumulative node strength distribution for the directors network.
It appears that the strength distribution is closely related to the degree distribution, see Fig.~\ref{dcddf}.
Indeed, the average strength $\langle s(k) \rangle$ of nodes with degree $k$, when there is no particular influence of topology on weights, simply follows an expected linear behaviour $\langle s(k) \rangle = \langle w \rangle k$, where $\langle w \rangle$ is the average weight of a link. In our case, the linear fit is almost perfect, with an adjusted r-squared correlation coefficient of $0.9955$; in fact, a power-law fit of the same model would yield an even higher r-squared of $0.9988$, but the exponent would be $1.014 \pm 0.013$. The same holds for the boards projections.
This linear growth of the average strength as a function of vertex degree, agrees with the findings on other networks resulting from the projection of bipartite graphs, such as scientific collaboration networks~\cite{bart05}.


\begin{figure}[!h] \center
 \includegraphics[width=0.4\textwidth]{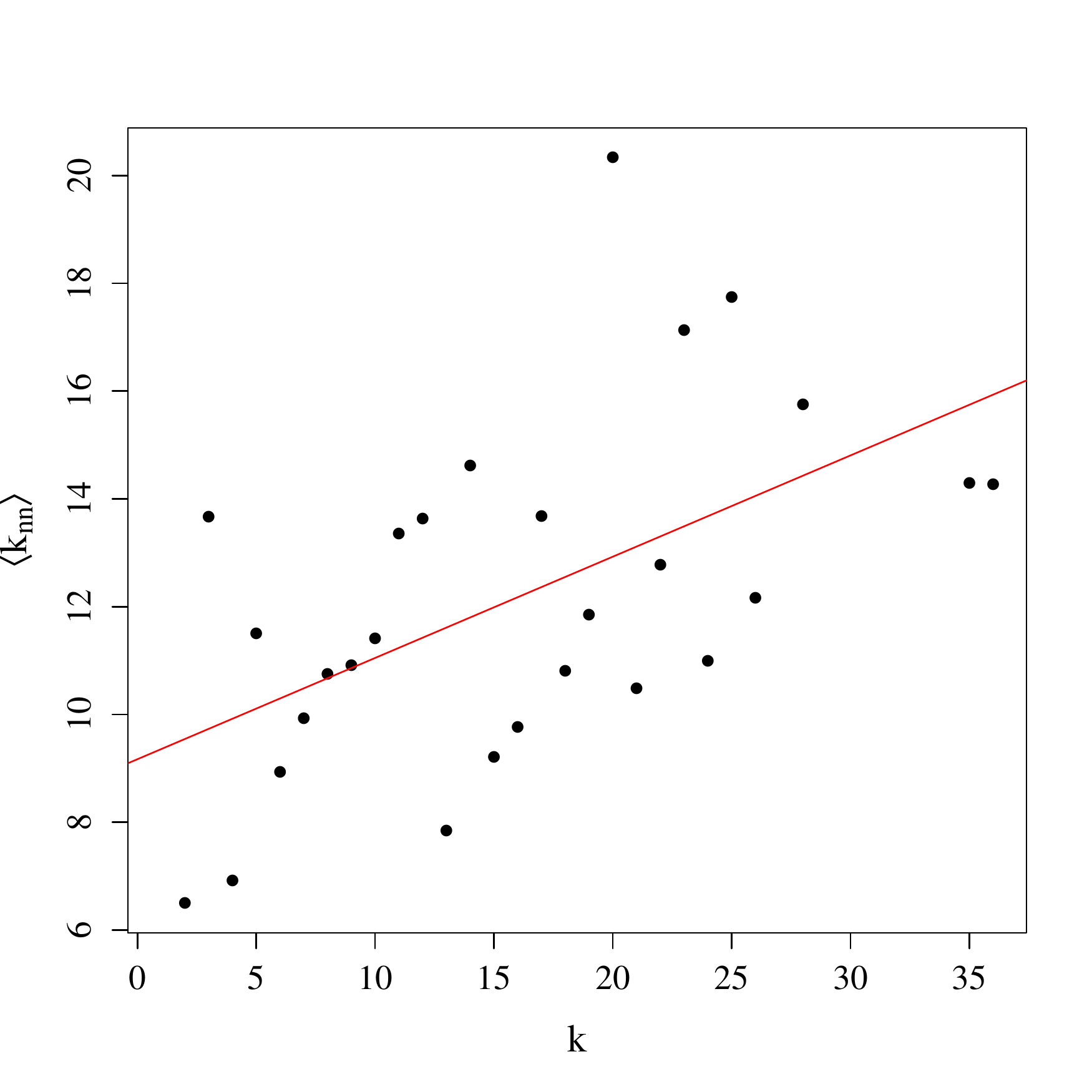}
 \caption{Average degree of nearest neighbors as a function of vertex degree in the largest connected component of the directors projection. Red line shows linear regression (slope $0.18792$, p-value $0.00293$).}
 \label{knn}
\end{figure}

Figure~\ref{knn} is a plot of the average mean degree $\langle k_{nn} \rangle$ of the neighbors of the nodes with degree $k$ in the case of the directors
graph. This provides an easy to compute
approximation to de\-gree-de\-gree correlation~\cite{satorras-corr} and shows an assortative behavior, i.e. high degree vertices tend
to have high degree neighbors. This seems to be a general feature of real social networks~\cite{newman-03,newman-book} and has
been found in other directors networks~\cite{calda1,battiston1}. The same plot for the boards networks is not statistically significant
in our case due to insufficient data.

\begin{table*}[hptb]
\begin{center}
\caption{Ranking of the $10$ most central directors with respect to the the four centrality measures evaluated on the directors projection.} 
\label{centrality-dir}
\begin{tabular}{ccccc} \toprule
\multicolumn{1}{c}{Rank}&\multicolumn{1}{c}{Degree Centrality}&\multicolumn{1}{c}{Eigenvector Centrality}&\multicolumn{1}{c}{Betweenness Centrality}&\multicolumn{1}{c}{Closeness Centrality}\\
\cmidrule{1-5}
$1^{st}$ &Peter Brabeck-Letmathe &Herbert Bolliger &Daniel J.Sauter &Daniel J.Sauter\\
$2^{nd}$ &Herbert Bolliger &Beat Zahnd &Fritz Studer &Peter Kuepfer\\
$3^{rd}$ &Theo Siegert &Oswald Kessler &Doris Russi Schurter &Monika Ribar\\
$4^{th}$ &Paola Ghillani &Paola Ghillani &Peter Kuepfer &Peter Brabeck-Letmathe\\
$5^{th}$ &Pius Baschera &Ernst Weber &Urs Widmer &Daniel Borel\\
$6^{th}$ &Ernst Tanner &Andrea Broggini &Peter Brabeck-Letmathe &Fritz Studer\\
$7^{th}$ &Beat Zahnd &Christian Biland &Conrad Löffel &Dieter Spälti\\
$8^{th}$ &Oswald Kessler &Claude Hauser &Monika Ribar &Andreas von Planta\\
$9^{th}$ &Ulrich Gygi &Doris Aebi &Daniel Borel &Rolf P.Jetzer\\
$10^{th}$ &Andreas Koopmann &Fabrice Zumbrunnen &Paola Ghillani &Charles G.T Stonehill\\
\bottomrule
\end{tabular}
\end{center}
\end{table*}

\begin{figure*}[ht!]\center
 \includegraphics[width=0.75\textwidth]{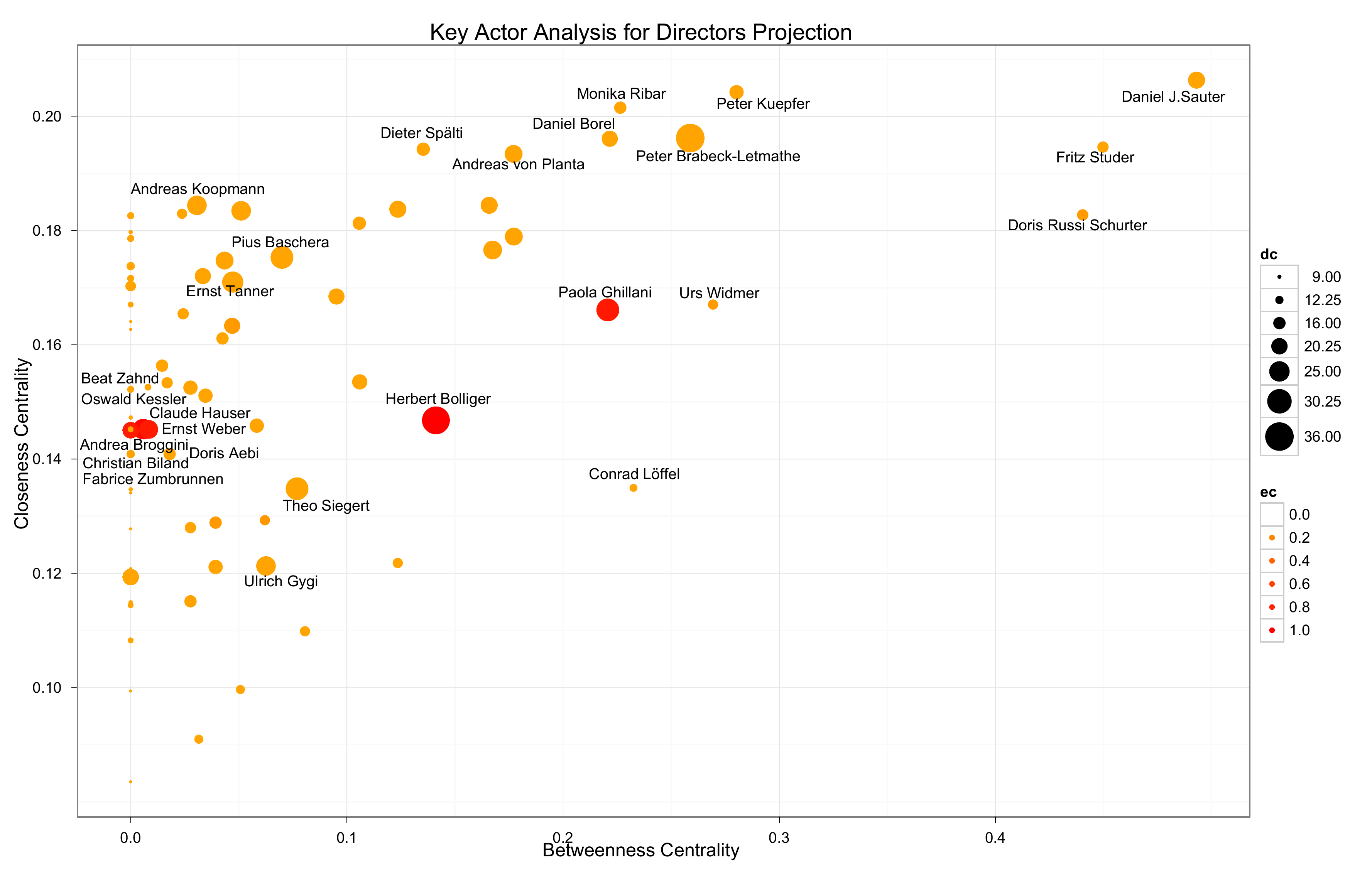}
 \caption{Directors network: key-actors analysis. Each point in the plane corresponds to a director; x and y coordinates define its betweenness and closeness centrality scores, respectively. Point size grows with degree (dc), color gets darker with eigenvector centrality (ec) score (see legend
 on the side). Due to readability reasons, only those who rank higher are labeled (refer to Table~\ref{centrality-dir}).}
 \label{key-directors}
\end{figure*}

\subsection{Centrality}

In any social network it is of importance to try to assess which actors play a key role with respect to
the rest of the network. This idea can be made quantitative through the use of \textit{centrality} measures.
Here we have used the \textit{betweenness centrality}, the \textit{closeness centrality}, the
\textit{eigenvector centrality} and the \textit{degree centrality}~\cite{Kola}.

Degree centrality is a straightforward  measure that simply attributes more importance to highly connected actors.
However, it is local in character and does not take into account the global network environment. Contrastingly,
the other three measures are more informative as they take into account the whole structure of the graph
in different ways in evaluating the centrality of a node. 
The betweenness $b_v$ of a node $v \in V$ is defined as: 

$$ b_v = \sum_{i \ne v \ne j} \frac {n_{ij}(v)} {n_{ij}}
$$

\noindent where $n_{ij}$ is the total number of shortest paths between $i$ and $j$, and $n_{ij}(v)$ is the number of those shortest
paths that go through $v$. Nodes with high betweenness are more central in the sense that they have more control since more
traffic goes through them. Nodes with high betweenness play the role of ``brokers'' in a social sense.

\begin{table*}[hptb]
\begin{center}
\caption{Ranking of the $10$ most central companies with respect to the four centrality measures evaluated on the boards projection.} 
\label{centrality-boards}
\begin{tabular}{ccccc} \toprule
\multicolumn{1}{c}{Rank}&\multicolumn{1}{c}{Degree Centrality}&\multicolumn{1}{c}{Eigenvector Centrality}&\multicolumn{1}{c}{Betweenness Centrality}&\multicolumn{1}{c}{Closeness Centrality}\\
\cmidrule{1-5}
$1^{st}$ &Credit Suisse &Denner &Julius Baer &Julius Baer\\
$2^{nd}$ &Rieter &Migros &Helvetia &Holcim\\
$3^{rd}$ &Roche &Hotelplan &Sika &Sika\\
$4^{th}$ &CFF &Banque Migros &Luzerner KantonalBank &Logitech\\
$5^{th}$ &Clariant &Migrol &Holcim &Nestl\'e\\
$6^{th}$ &Holcim &Helvetia &Barry Callebaut &Luzerner KantonalBank\\
$7^{th}$ &Hotelplan &Kuehne-Nagel &Roche &Rieter\\
$8^{th}$ &Julius Baer &Raiffeisen &Valora &Sulzer\\
$9^{th}$ &Migros &Barry Callebaut &CFF &Novartis\\
$10^{th}$ &Banque Migros &Luzerner KantonalBank &Syngenta &Roche\\
\bottomrule
\end{tabular}
\end{center}
\end{table*}

\begin{figure*}[!ht]\center
 \includegraphics[width=0.75\textwidth]{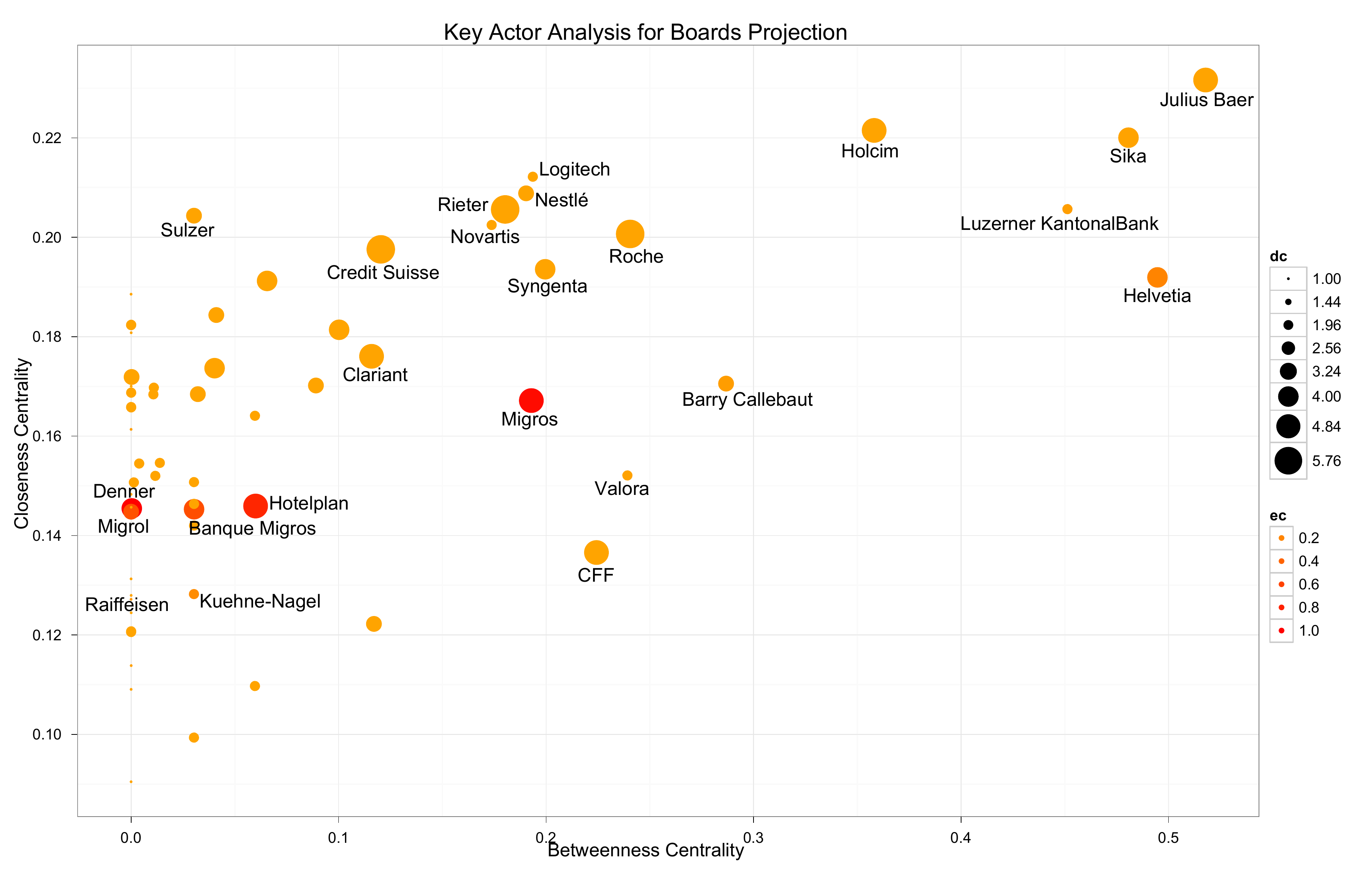}
 \caption{Boards network: key-actors analysis. Each point in the plane corresponds to a company; x and y coordinates define its betweenness and closeness centrality scores, respectively. Point size grows with degree (dc), color gets darker with eigenvector centrality (ec) score (see legend
 on the side). Due to readability reasons, only those who rank higher are labeled (refer to Table~\ref{centrality-boards}).}
 \label{key-boards}
\end{figure*}

Closeness centrality gives the average distance of a given
node to all others and is expressed by the following formula:

$$ h_i = \frac {1} {\sum_{k \ne i} l_{ik}}$$

\noindent where $l_{ik}$ is the shortest path from node $i$ to node $k$. As a consequence, nodes that have small
shortest paths distances to other nodes will enjoy high centrality under this measure since they are
``closer'' to the other nodes.

The third centrality measure is due to Bonacich~\cite{bonacich1987power}, according to whom the centrality of
a node depends on how central are its neighbors and it can be expressed in terms of the eigenvectors
of the adjacency matrix $A(G)$ of graph $G$. 
Table~\ref{centrality-dir} gives a summary of the above described measures for the top $10$  actors in the directors
network and Table~\ref{centrality-boards} does the same for the boards.

The various centrality measures are not necessarily correlated among them, as can already be spotted in the tables. For this reason,
we offer in Fig.~\ref{key-directors} a global view of the correlation between the four centrality measures for the same best-ranked actors. 
The placement of points in the plane reflects
potential correlation between closeness and betweenness, whereas the circles' area stands for degree, and circles' color represents
eigenvector centrality. The directors that find themselves on the top right corner are very central in the network in the sense
of the paths leading or passing through them, as directors ``Daniel J. Sauter'', ``Fritz Studer'', and ``Peter Kuepfer''. However, they are 
not necessarily the most well connected ones, as shown by the
relatively small size of the corresponding circles and their light colors. On the other hand,
some directors are both highly connected locally and have neighbors that are also well connected, which is reflected by the size
and darkness of their circles, but in general they don't score very high in terms of betweenness. Directors ``Paola Ghillani'',  ``Peter 
Brabeck-Letmathe'', and ``Herbert Bollinger'' seem to enjoy a high amount of centrality in the network, as they rank high with
respect to at least three measures.

An equivalent investigation can be performed on the boards projection by looking at Table~\ref{centrality-boards} and Figure~\ref{key-boards}.
In this case the private bank ``Julius Bär'', which ranks $8^{th}$ as for degree, has the combined highest betweenness and closeness centrality. 
 We didn't perform a complete cross-correlation
among the scores in the two projections, nevertheless we observe that director Daniel J. Sauter, who has an identical position in the plot of Fig.~\ref{key-directors},
is actually a board member of Julius Bär, and of ``Sika'' as well. Similarly, director Fritz Studer sits in both Sika and in the ``Luzerner KantonalBank'', with the
latter ranking among the top $10$ in three of four measures despite its small degree. Focusing on eigenvector centrality alone, it is the ``Migros''
Group that stands out with its associate companies, which are tightly coupled among themselves. Interestingly again, central directors as Paola Ghillani and Herbert Bollinger sit on those boards.
We refrain from moving the discussion to the point of view of management and organization, however, the analysis of several centrality measures at a time and on the two projections, permits to highlight those actors who play a key role in the topology of the network.



\begin{figure}[!htb]
 \includegraphics[width=0.5\textwidth]{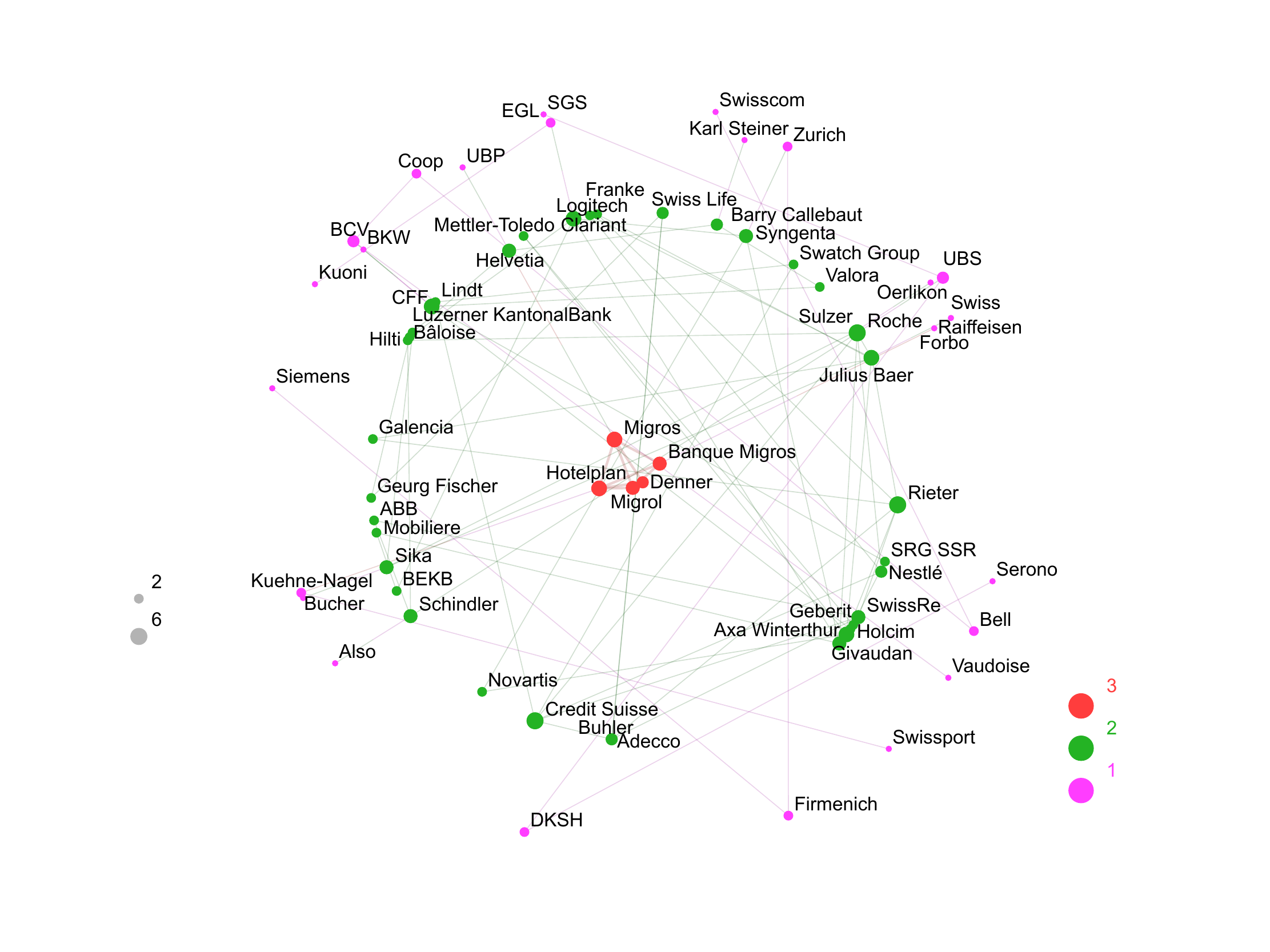}
 \caption{Boards network: $k-$core decomposition of the largest connected component. Nodes size scales with vertex degree, nodes color depends on the shell index (see lateral legends).}
 \label{Bcores}
\end{figure}

\subsection{Shells and Communities}

Large networks are typically difficult to visualize in two dimensions but
any network can be depicted according to its \textit{k-core decomposition}, which facilitates the graph layout. 
The k-core of a graph $G$ is the connected maximal induced subgraph which has minimum degree
greater than or equal to $k$~\cite{k-core1}.
This decomposition can be obtained
by a recursive pruning of the least connected vertices and it allows to disentangle the hierarchical structure
of networks by progressively focusing on their central cores. Figure~\ref{Bcores} shows such a representation for the boards network obtained with
the  \textit{LaNet-Vi} software~\cite{alvarez2005k}. There is a clear central $3$-core shell formed by the following firms: ``Migros'', ``Denner'',
``Migros Bank'', ``Hotelplan'', and ``Migrol''. The existence of this shell is not surprising, given that the component firms are
all associates of Migros but it is interesting to point it out since it is not immediately apparent without performing
the core analysis. From a board management point of view it is quite reasonable that some directors might be
shared among companies belonging to the same group. We also remark that coreness and degree are not
necessarily correlated: larger values of coreness correspond to nodes with both larger degree and more central
position in the network’s structure.
This can be seen in Table~\ref{centrality-boards} where firms belonging to the Migros group happen to possess high
eigenvector centrality and, for some of them also high degree. However, ``Credit Suisse'', which has the highest
degree but is not otherwise very globally central, belongs to the $2$-core.
In fact other banks like ``UBS'' and ``BCV'', lie in the outer shell despite having a degree of $3$, which is above average. 
In conclusion, the $k$-core view of the boards network shows that there is not a clear global hierarchical
structure, nonetheless, with the aid of such a tool, the network fingerprinting is quite clear.

%
%
%
%

\begin{figure}[h!]
\begin{center}
 \includegraphics[width=0.5\textwidth]{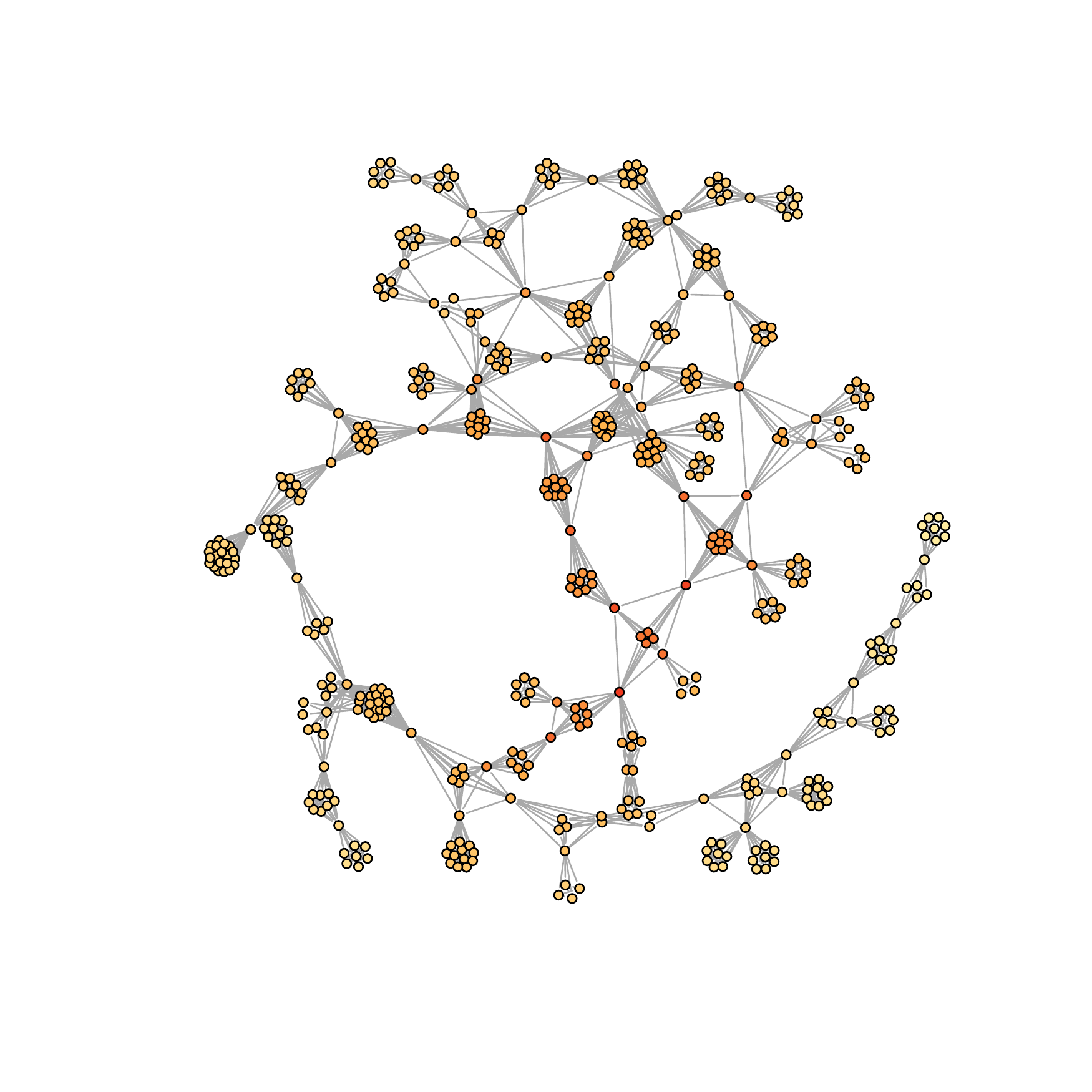}
 \caption{Largest connected component of the directors network. Nodes color scales with the closeness centrality score of the corresponding vertex: the darker the color, the higher the ``efficiency'' in global information spreading when starting from the considered point.}
\label{dir-netw}
\end{center}
\end{figure}

It is less so for the projection on the directors' set. That network is mainly composed by cliques (the boards) which are loosely
connected among themselves (through the interlocks). This would induce a $k$-core decomposition in which a simple concentric arrangement 
by shell index would not be possible, because some cores would present separated components, i.e. cliques having the same coreness
value but no connections with the rest of their $k$-shell. For this reason we argue that such a representation would not help to reveal the
structure of the network, to the contrary, it would make it more confused. This is why
the k-core decomposition for the giant connected component of the directors network is not shown.
The whole picture is given instead in Figure~\ref{dir-netw}, by means of a more standard force-based layout algorithm.


\subsection{Tie Importance and Information Spreading}

In the same way as viruses can spread from a person to another in a network of people contacts, or from a computer to another 
in the case of computer viruses, ideas can spread through a social network in a kind of
contagion process. Actually, there are differences between the two cases, for viruses can spread with a certain probability
related to their infectiousness and the state of the target person or computer without any clear decision-making process
of the latter, whereas
in social contagion agents may evaluate ideas and decide whether to accept them or not and are submitted to other
external influences as well, such as broadcasting, existing practices, and the media in general. However, when the people's
decision processes are unknown, or difficult to model, a random model similar to the ones used in epidemiology is a
useful starting point. In this spirit, we present in the following a numerical analysis of information spreading through the
board directors network. This analysis may shed some light on the influence of the directors' network on the way information
travels among people and thus among the boards themselves.

The model is a very simple SI (susceptible-infected) model in which  there are only two states: either a node is susceptible
or it is infected. A susceptible node may become infected with a certain probability if it has a neighbor who is in the infected
state. When a node has been infected, it remains infected forever. Let us suppose that node $i$ is in the infected state at time
$t$; then,
if node $j \in V_i$, $j$ becomes infected at time $t+1$ with probability $ \beta \times w_{ij}$, where  $\beta$ is the so-called
infection rate and $w_{ij}$ is the link weight. In the present case, in which no disease is implied,  $\beta$ might represent 
an unknown average speed with
which information such as news and rumors flows through neighbors in the network. Similar models, in which the transmission rate
also depends on the link weight, have been used in
an economical setting in~\cite{Antonios-Econ} and in an actual large social network in~\cite{OnnelaPNAS}.

\begin{figure}[!h]\center
 \includegraphics[width=0.4\textwidth]{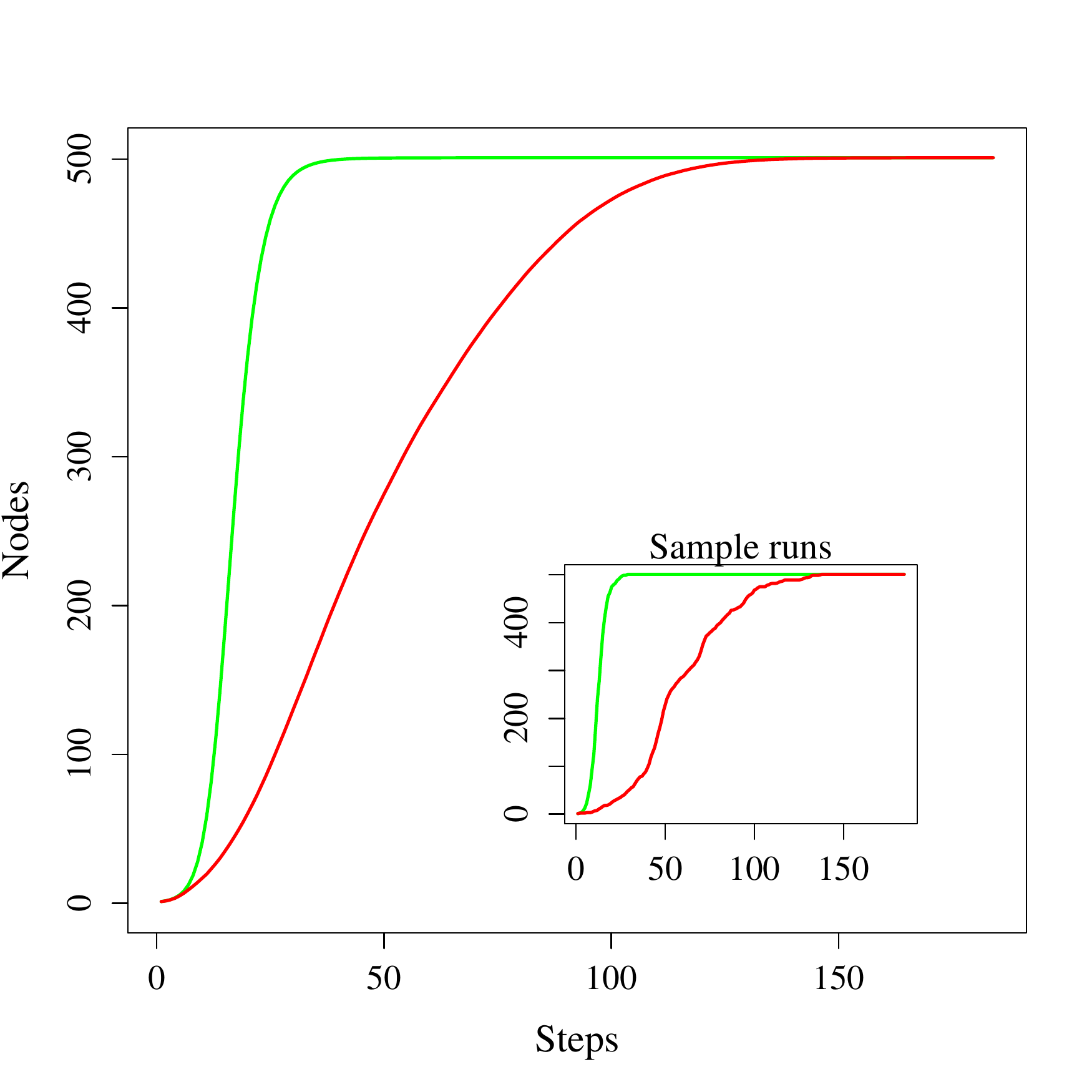}
 \includegraphics[width=0.4\textwidth]{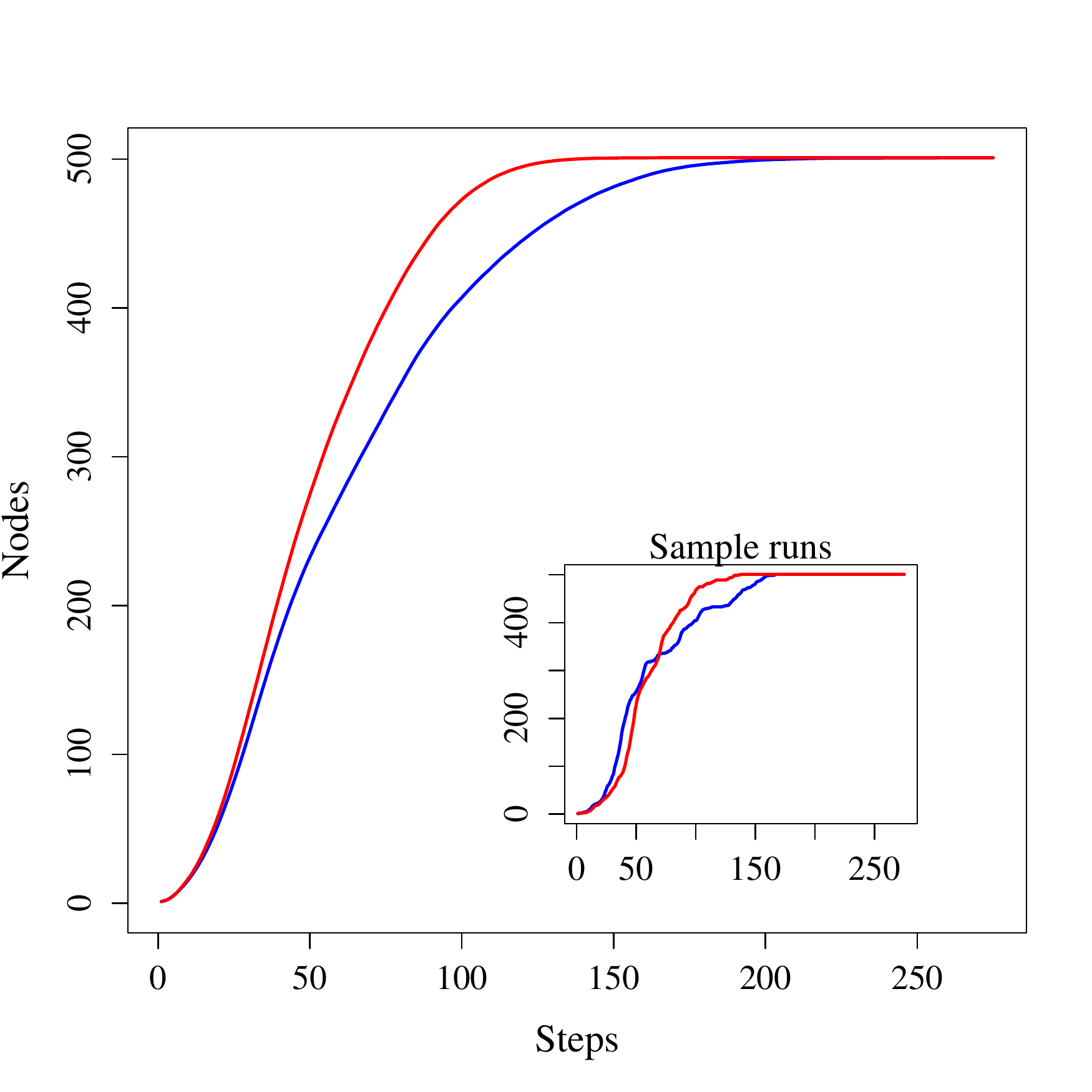}
 \caption{Number of infected nodes as a function of time in the giant connected component of the directors network.
 Curves are averages over $500$ runs. Top image: the red curve refers to the real network, while the
 green one corresponds to a family of graphs obtained by randomizing the real graph but keeping the
 degree sequence unchanged. The inset shows two single executions. Bottom image: the red curve is
 the same as in the top image; the blue curve corresponds to the real network after suppressing
 $7\%$ of the links in decreasing betweenness order. The inset shows two particular runs. Note that
  in the top and bottom plots the horizontal scales differ.}
 \label{Epidemics}
\end{figure}

We ran $500$ simulations of the epidemics process in the giant connected component of the 
directors graph starting each time from a single randomly chosen infected node
with $\beta=0.05$. The $\beta$ parameter has only a scaling role in the process, influencing the
rate at which information travels through the network but it doesn't change the relative behaviors on different topologies.
The average results  are shown in Figure~\ref{Epidemics} (top image, red curve), in which are also reported (green curve) the results
 corresponding to the same
process carried out on a family of randomized networks having the same size and degree sequence as the original directors network.
An interesting effect is immediately apparent from the figure: the propagation speed in the real network is notably slower than 
on the randomized versions. This would not be surprising if the resulting random graphs were of the Erd\"os-R\'enyi  type~\cite{bollobas}.
However,
in randomizing, we kept the degree sequence invariant, which means that the resulting network, though random,
does not have a Poisson degree distribution. Thus, the slowing down must be caused by particular topological features
of the directors network, which is depicted in Figure~\ref{dir-netw}.

We attribute the different shape of the information propagation 
(see inset of the top image of Fig.~\ref{Epidemics}), which is less smooth in the real network compared to the randomized one,
to the presence of well connected clusters of directors, almost cliques, that are clearly visible in Fig.~\ref{dir-netw}. 
Because of this, the information flow has first to propagate within a cluster before being able to conquer another one. This interpretation is
confirmed by the bottom image of Fig.~\ref{Epidemics}, in which the red curve is the same as before, whilst the blue one
represents the average propagation speed with a fraction $0.07$ of the highest betweenness links being suppressed, which
corresponds to a loss of only $19$ edges out of $2546$.
From our computations, and as it can also be noticed from Fig.~\ref{dir-netw}, in this network there are no bridges, i.e. links whose cutting would cause the graph to
fall apart into two separated connected components, neither local-bridges, i.e. edges whose endpoints have
disjoint sets of nearest neighbors. Thus, after
the suppression of these links, the graph remains connected but, although the number of links being suppressed is
very small with respect to the total, the global information spreading is significantly slowed down. This can be explained by the
fact that, although clusters are not affected, there are less paths available among them. Indeed, looking at the inset in
the right figure, which shows a single run, the inter-cluster communication is even less smooth and takes a little more time when
 such links are cut (blue curve in the inset of the bottom figure).

We are aware that the previous information diffusion model is, at best, only a very rough approximation of
the actual human processes that might take place between directors and their boards in real-life. 
It is difficult to relate the results found for a highly idealized
model of information spreading with actual decision processes in the boards, where people talk, vote, and submit ideas
through complex and largely unknown communication and decision patterns. In this context,  network topology is only a rough
proxy for aggregating all these rich human interactions. 
With such a caveat in mind,
the conclusions of this section can be summarized as follows. The particular structure of this directors network has a marked
influence on the way in which information flows through the network. The presence of small densely connected clusters, which
are typical of these kind of projections of bipartite affiliation networks, have the effect of slowing down the epidemic process with
respect to randomized versions of the same network. What could be said is that opinions, practices, and ideas, will have more 
time to mature and evolve in a such a network structure than in an arbitrary one.


\section{Conclusion}

Starting from empirical data defining a bipartite graph in which a set of vertices, the directors, have links with
another set, the boards, when a director sits in a given board, we have produced two projection graphs:
the directors graph and the boards graph for the Swiss top $108$ companies in 2009. This is an interesting case study
because it deals with the main companies' organization in an economically and financially important country during the present
 crisis. 

First we have studied
a number of standard statistics of these graphs: average degree, degree distribution, weight and strength distribution,
and degree-degree correlation. The results of these measures are in general comparable with those 
of the few preceding similar studies~\cite{davisUS,calda1,battiston1} dealing with US and Italian boards, with some exceptions.
The main differences are the smaller size of the whole networks, as well as of their giant connected components, in the Swiss case, which is 
related to the smaller size of the country itself, and the smaller number of interlocks. 
Besides the basic statistical study, we have performed new investigations with the goal of highlighting the key actors
and connections in the networks. To this purpose we used several centrality measures and their correlations, and a $k$-core analysis of the
whole networks. This study has allowed us to find out a few directors and boards that play a central role in the topological
sense. Of course, we are careful not to draw any conclusions about management or governance from these statistics, but we
highlight a potential role of these actors in the networks.

Finally, we set out to a study of the way in which information, such as rumors, ideas, or practices, may spread
through the directors network. Using a very basic SI model and computer simulations, we showed that the particular
structure of the directors network strongly influences the way in which information flows. Indeed, the average propagation
speed is notably slower in the real network than in randomized versions of it. It appears that the many cluster structures
present in the directors network are responsible for the slowing down. This has been confirmed by a second simulation study
in which a small fraction of the most central links is removed. The result is that the spread is further slowed-down, and thus
these links have an important role in inter-cluster communication.

As future works we think that it would be important to complement the present study with an investigation of the Swiss boards
in years before 2007 in such a way that the evolution of the networks in this crucial time frame be evaluated. This requires
a time-consuming data gathering activity but the results might prove useful for a better understanding how the economical
system of the country has reacted to the crisis at the level of board governance at least. Another interesting study would be
a comparison of the Swiss systems with other European countries, for which small samples  are available~\cite{heidrick-report},
by completing that data sets and performing a cross-comparison study of the network aspects.

\bibliographystyle{elsarticle-num}
\bibliography{boards}

\end{document}